\documentclass{iopjournal}

\usepackage{amsmath}
\usepackage{enumitem,amssymb}
\usepackage[normalem]{ulem}
\usepackage{cite} 
\usepackage{booktabs}

\begin{document}

\articletype{Paper}

\title{Improved Ising Meson Spectroscopy Simulation on a Noisy Digital Quantum Device }

\author{Hao-Ti~Hung$^{1,2,*}$\orcid{0009-0002-6977-0794}, Isabel Nha Minh Le$^{3,4}$\orcid{0000-0001-6707-044X}, Johannes Knolle$^{5,4,6}$\orcid{0000-0002-0956-2419} and Ying-Jer~Kao$^{1,2,*}$\orcid{0000-0002-3329-6018}}

\affil{$^1$Department of Physics, National Taiwan University, Taipei 10617, Taiwan}

\affil{$^2$Center for Theoretical Physics, National Taiwan University, Taipei 10617, Taiwan}

\affil{$^3$Technical University of Munich, School of Computation, Information and Technology, Boltzmannstraße 3, 85748 Garching, Germany}

\affil{$^4$Munich Center for Quantum Science and Technology (MCQST), Schellingstrasse 4, 80799 Munich, Germany}

\affil{$^5$Technical University of Munich, TUM School of Natural Sciences,
Physics Department, TQM, 85748 Garching, Germany}

\affil{$^6$Blackett Laboratory, Imperial College London, London SW7 2AZ, United Kingdom}

\affil{$^*$Author to whom any correspondence should be addressed.}

\email{hunghaoti852@gmail.com and yjkao@phys.ntu.edu.tw}


\begin{abstract}
The transverse-field Ising model serves as a paradigm for studying confinement and excitation spectra, particularly the emergence of $E_8$ symmetry near criticality. However, experimentally resolving the Ising meson spectroscopy required to verify these symmetries is challenging on near-term quantum hardware due to the depth of circuits required for real-time evolution. Here, we demonstrate improved spectroscopy of confined excitations using two distinct error-resilient circuit construction techniques on the IBM Torino device: first-order Trotter decomposition utilizing native fractional gates, and a tensor-network-based circuit compression via Riemannian optimization. By analyzing the Fourier spectrum of error-mitigated time-series data, we successfully identify key signatures of $E_8$ symmetry despite hardware noise. These results validate the viability of both circuit compression and hardware-efficient compilation for probing complex topological phenomena on NISQ devices.
\end{abstract}

\section{Introduction}
\label{sec:intro}

The advent of quantum devices in the  Noisy Intermediate-Scale Quantum (NISQ) era~\cite{Preskill_2018} allows us to perform quantum simulations even in the absence of quantum error correction. The simplest manybody quantum model, the one-dimensional transverse field Ising model (TFIM), offers a minimal setting to study quantum phase {transitions~\cite{Pfeuty_1970,Sachdev_2011}}. When the transverse field is tuned to a critical value, the system undergoes a phase transition, and surprisingly, its low-energy behavior reveals a hidden  structure. 

The TFIM with a longitudinal field can be described by the  Hamiltonian, 
\begin{equation}\label{eq:Ising_H}
    H = -\left(\sum_i \sigma_i^z \sigma_{i+1}^z + h_x \sum_i \sigma_i^x + h_z \sum_i \sigma_i^z\right),
\end{equation}
where \(h_x\) and \(h_z\) are transverse and longitudinal magnetic fields, respectively. In the absence of a longitudinal field ($h_z = 0$), the model maps to free fermions, and for small transverse field $h_x$, its low-energy excitations consist of pairs of kinks (domain walls) separating regions of opposite magnetization. These domain walls can propagate freely without a longitudinal field. In contrast, introducing a finite longitudinal field ($h_z > 0$) generates a linear confining potential between domain walls, resulting in bound states that exhibit meson-like behavior~\cite{Rutkevich_2008, Kormos_2017, Robertson_2025}. This allows us to study the dynamics of these confined excitations by simulating Eq.~\eqref{eq:Ising_H} at the critical point. 

In the scaling limit, the continuum field theory describing this regime maps to the integrable $E_8$ model~\cite{ZAMOLODCHIKOV_1989, Delfino_1995}, which supports eight stable particles (mesons) whose mass ratios are exactly fixed by the structure of the $E_8$ Lie algebra, and the ratios of the eight meson masses take universal values. This $E_8$ spectrum has been investigated extensively using tensor network (TN) methods~\cite{Jonas_2011,Kormos_2017,Mazza_2019,Castro-Alvaredo_2020,Ramos_2020,Luka_2025}, and has also been observed experimentally in quasi-one-dimensional Ising-like magnets, including the ferromagnetic material \(\mathrm{CoNb_2O_6}\)~\cite{Coldea_2010} and the antiferromagnetic material \(\mathrm{BaCo_2V_2O_8}\)~\cite{Zhang_2020,Zou_2021}. However, simulations of this physics on quantum hardware only occur recently~\cite{Vovrosh_2021, Lamb_2024}, due to  challenges posed by  noise and circuit depth limitations in current devices. Moreover, accurately extracting the meson spectrum from real-time dynamics is particularly demanding, as resolution in the low frequency data in the Fourier spectrum requires simulation of long-time dynamics. 

The goal of this work is to simulate this physics using  IBM’s latest  quantum device 
\texttt{ibm\_torino} to  extract the meson masses from the real-time dynamics.  We compare the results with the $E_8$ predictions to assess to what extent the symmetry can be reproduced~\cite{Vovrosh_2021}.
However, the required deep circuit depth to perform a long real-time evolution poses significant challenges to previous simulations of  TFIM dynamics~\cite{Hebenstreit_2017,CerveraLierta_2018,Zhukov_2018,Smith_2019,Vovrosh_2021, Mueller_2023, Lamb_2024, Vovrosh_2025}.
To mitigate the limitations due to noise and decoherence, classical techniques such as TN~\cite{Lin_2021, Astrakhantsev_2023, Robertson_2025} have been proposed to compress  quantum circuits to reduce the circuit depth while maintaining accuracy. In addition, the Heron-class QPU inside \texttt{ibm\_torino}  supports native fractional gates, allowing for direct gate operations \texttt{RZZ}$(\theta)$  for $ 0<\theta \leq \pi / 2 $ and \texttt{RX}$(\theta)$  for any $\theta$, which significantly reduce the gate error. 

Here we explore two different methods for constructing the time-evolution operator: The first approach uses first-order Trotter decomposition and {constructs} the real-time evolution operator directly using native gate operations provided on the IBM device  \texttt{ibm\_torino}. The second approach employs Riemannian optimization~\cite{Le_2025,kotil2024riemannian,Putterer2025riemannian}, a TN-based technique that approximates the evolution operator using a fixed-depth variational circuit. Using  an error mitigation strategy to correct for decay in the measured observables, we show that good agreement with the mass excitations predicted by the $E_8$ symmetry can be reached.

The remainder of this article is organized as follows. Section~\ref{sec:E8} introduces the $E_8$ symmetry in the 1D Ising model with transverse and longitudinal fields. Section~\ref{sec:IBMQ_real} explains quantum circuit construction and error mitigation. Section~\ref{sec:TN_Compress} describes the use of Riemannian optimization for low-depth circuit compression. Section~\ref{sec:Freq_domain} presents simulation results on IBM Quantum devices in the frequency domain and compares them with theoretical predictions. Finally, we conclude in Sec.~\ref{sec:conclusion}

\section{\texorpdfstring{$E_8$}{E8} symmetry in quantum Ising model }
\label{sec:E8}

The 1D TFIM,  at the quantum critical point (\(h_x = 1, h_z = 0\)), can be described by the Ising conformal field theory (CFT) with central charge \(c = 1/2\). When the critical theory is perturbed by a small longitudinal field (\(h_z \neq 0\)), it relates to {the $E_8$ Lie-algebra} symmetry {at} the scaling limit~\cite{ZAMOLODCHIKOV_1989, Delfino_1995}, where the relationship between meson masses can be excatly determined (Tab.~\ref{tab:e8_mass_spectrum}).

To perform the quantum circuit simulations, we prepare an initial state with two domain walls, such that the spin points up at the boundaries and down in the center (For example, \(|\uparrow \uparrow\uparrow \downarrow \downarrow \uparrow\uparrow\uparrow\rangle\) in an 8-site chain.) 
The dynamical structure factor, $S^y(k,\omega)$, is a good quantity to  extract meson masses from spin systems, which can be found by a two-point correlation function, $C^y(i,t)=\langle\psi_0|\sigma^y_i(t)\sigma^y_i(0)|\psi_0\rangle$~\cite{Coldea_2010, Jonas_2011,Wu_2014,Ramos_2020}. However,  simulating this correlator on NISQ devices remains difficult. Instead, several studies have shown that the meson spectrum can be extracted from the entanglement spectrum~\cite{Castro-Alvaredo_2020}, or the dynamics of the total magnetization~\cite{Kormos_2017,Liu_2019,Tindall_2024,Luka_2025},
$
\langle \sigma^z(t) \rangle = \frac{1}{L} \sum_{i=1}^L \langle \sigma^z_i(t) \rangle
$.
Here, we propose that  the  central-site magnetization,
$
\langle \sigma^z_{\mathrm{cen}}(t) \rangle,
$
provides enough information to extract the spectrum from the frequency domain, \(\langle \sigma^z_{\mathrm{cen}}(\omega) \rangle\).
These spectral peaks are expected to match the mass ratios predicted by the $E_8$  quantum field theory.

\begin{figure*}[tbp]
  \includegraphics[width=\linewidth, trim = 0 0 0 0]
  {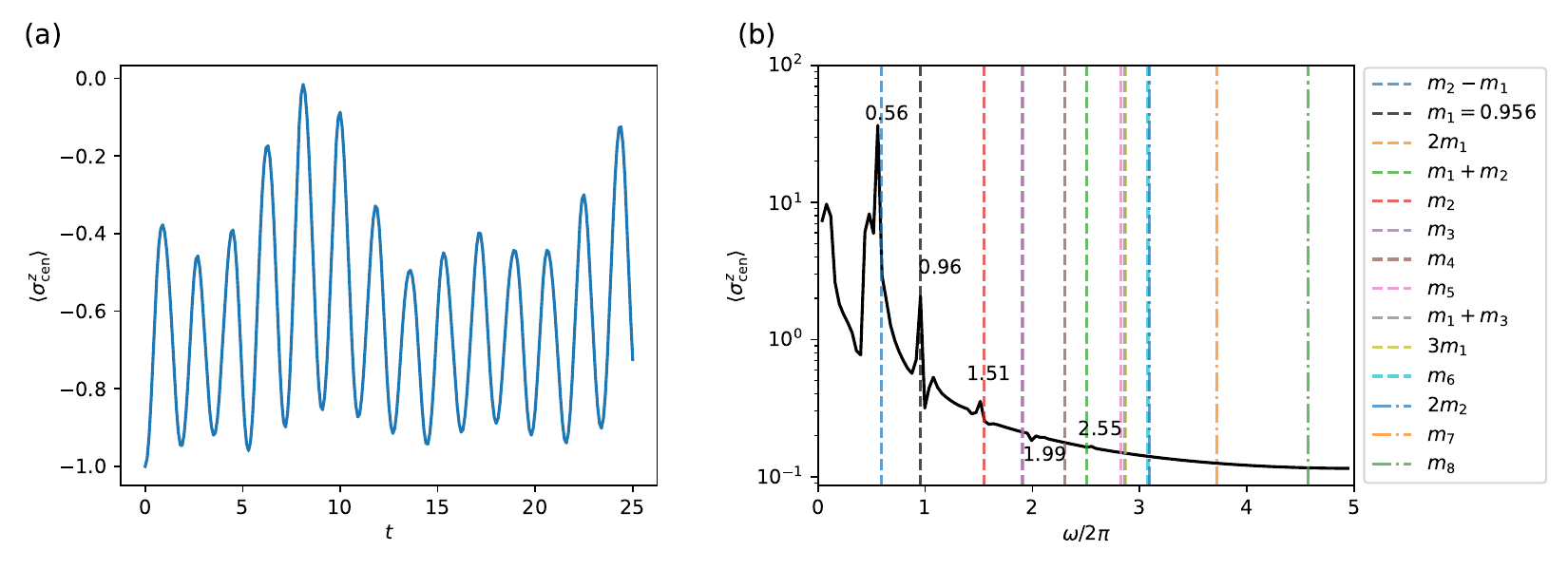}
  \caption{
Simulation results obtained via exact diagonalization on a classical computer. The system evolves from the initial state \(|\uparrow \uparrow \downarrow \downarrow \downarrow \downarrow \downarrow\downarrow\downarrow\uparrow \uparrow\rangle\) using the parameters \(h_x = 1\) and \(h_z = 3\). The observable is the central spin expectation value \(\langle \sigma^z_{\text{cen}}(t) \rangle\).  
(a) Time-domain dynamics of the observable.  
(b) Frequency-domain spectrum  plotted on a semi-log scale. Colored dashed lines indicate several expected peak positions predicted by the $E_8$ mass ratios. We identify the value of \(m_1\), and the positions of all other expected peaks are computed by applying the known $E_8$ mass ratios. The peak values obtained from ED are denoted directly on the plot.
}
  \label{fig:L11_E8}
\end{figure*}

To verify our approach, we carry out simulations on a classical computer to perform exact real-time evolution via exact diagonalization (ED) and compare the results with the expected $E_8$ spectrum. As shown in Fig.~\ref{fig:L11_E8}, we simulate a chain of 11 sites with the initial state prepared as \(|\uparrow \uparrow \downarrow \downarrow \downarrow \downarrow \downarrow \downarrow \downarrow \uparrow \uparrow\rangle\), with \(h_x = 1\) and \(h_z = 3\).~\footnote{{Although the $E_8$ spectrum formally emerges for a small perturbed longitudinal field, $h_z$, in practice a very small $h_z$ produces multiple closely spaced frequency components that are difficult to resolve in simulations. Increasing the longitudinal field enhances the separation and visibility of the meson peaks. For this reason, we choose \(h_z = 3\) to obtain clearer spectral signatures while still retaining the qualitative $E_8$ structure.}} The figure presents both the time-domain dynamics of \(\langle \sigma^z_\mathrm{cen}(t) \rangle\) at the center of the chain (Fig.~\ref{fig:L11_E8}(a)) and its corresponding frequency-domain representation obtained via Fourier transform (Fig.~\ref{fig:L11_E8}(b)). Clear spectral peaks are observed at frequencies that align with the theoretical $E_8$ mass ratios.

In addition to the exact real-time evolution, we simulate the dynamics using the time-dependent variational principle (TDVP)~\cite{Haegeman_2011,Haegeman_2016} within the TN framework, and also apply a first-order Trotter decomposition to construct the real-time evolution operator using TN methods, to mimic the statevector simulation used in quantum circuits. All simulations are implemented with the TN library \texttt{Cytnx}~\cite{Cytnx_2025}. Additionally, we benchmarked simulations across different system sizes \(L = 5, 8, 27\), and found that \(L = 8\) is sufficient to capture the essential spectral features relevant to our analysis. Further details and comparisons are presented in \ref{appdx:TN_res}.

\section{Simulations on IBM~Q}
\label{sec:IBMQ_real}
\subsection{Evolution gate}
\label{sub_sec:evo_gate}

\begin{figure*}[tbp]
  \includegraphics[width=\linewidth, trim = 0 0 0 0]
  {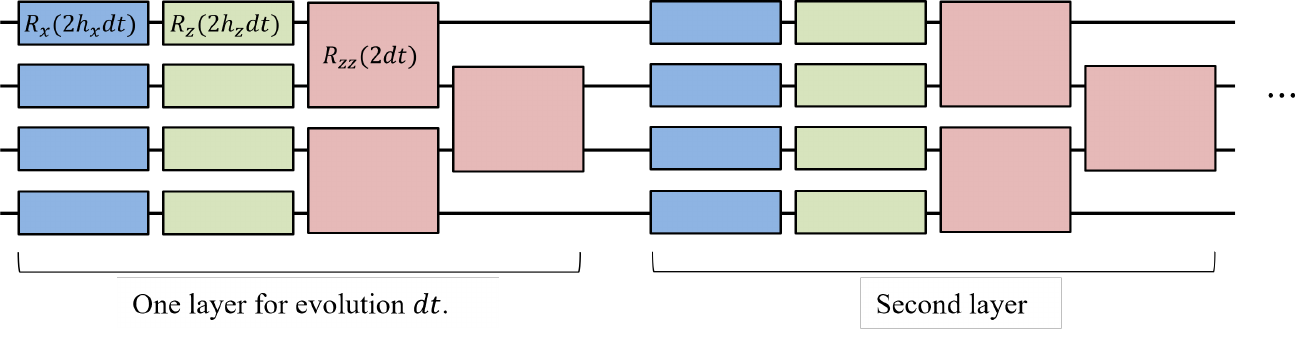}
  \caption{Quantum circuit construction of the real-time evolution operator for the TFIM using first-order Trotter decomposition, as described in Eq.~\eqref{eq:trotter}. The circuit consists of sequential layers implementing the longitudinal field (\texttt{RZ} gates), transverse field (\texttt{RX} gates), and Ising interaction (\texttt{RZZ} gates) terms. }
  \label{fig:trotter}
\end{figure*}

To perform real-time evolution on a quantum device, the first step is to construct the real-time evolution operator of the Hamiltonian. In the case of the TFIM with a longitudinal field (see Eq.~\eqref{eq:Ising_H}), we use the first-order Trotter-Suzuki decomposition and the time evolution operator over a short time interval \(dt\) can be approximated as
\begin{equation}\label{eq:trotter}
\begin{aligned}
U(dt) &= e^{-i dt H} \approx\ 
\left( \prod_i e^{i h_z dt\, \sigma_i^z} \right)
\left( \prod_i e^{i h_x dt\, \sigma_i^x} \right) \\
&\times 
\left( \prod_{\substack{i\in \text{odd}}} e^{i  dt\, \sigma_i^z \sigma_{i+1}^z} \right)
\left( \prod_{\substack{i\in \text{even}}} e^{i  dt\, \sigma_i^z \sigma_{i+1}^z} \right).
\end{aligned}
\end{equation}
Each exponential term corresponds to unitary gates that can be directly implemented using native quantum gates on IBM Q: \texttt{RZZ} for the interaction term and \texttt{RX}/\texttt{RZ} for the transverse and longitudinal fields. The quantum circuit representing Trotter step is shown in Fig.~\ref{fig:trotter}.

In our quantum simulations, we utilize the \texttt{Estimator} primitive provided from IBM Qskit~\cite{qiskit2024} to directly evaluate the time-dependent observable $\langle \sigma^z_{\text{cen}}(t) \rangle$, rather than constructing both the evolution operator $U$ and $U^\dagger$ around a measurement circuit. This approach allows us to access the expectation value of the central spin observable directly.

\subsection{Error mitigation}
\label{sub_sec:err_mit}
\begin{figure*}[htbp]
  \centering
  \includegraphics[width=\linewidth, trim = 0 0 0 0]
  {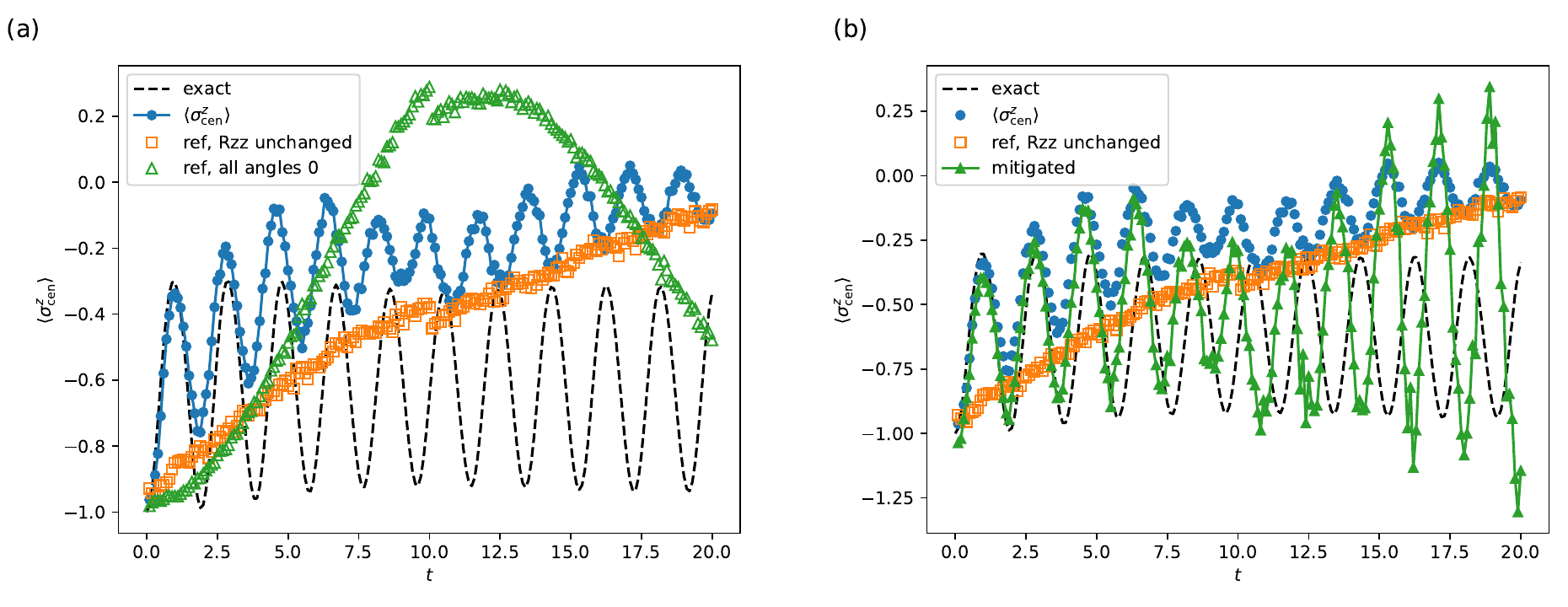}
  \caption{
Real-time evolution results on the IBM Quantum device \texttt{ibm\_torino} for an 8-site transverse-field Ising model with parameters \(h_x = 1\), \(h_z = 3\), initial state \(|\uparrow \uparrow \downarrow \downarrow \downarrow \downarrow \uparrow \uparrow\rangle\), and 8192 shots per circuit. The observable \(\langle \sigma^z(t) \rangle\) is measured at site 4 (the central site). (a) Comparison of two reference circuit strategies for error mitigation: \(\bullet\) markers correspond to the raw measured signal, \(\square\) markers correspond to setting all \texttt{RX} and \texttt{RZ} angles to zero while leaving \texttt{RZZ} gates unchanged, and \(\triangle\) markers correspond to setting all gate parameters to zero. A small discontinuity is visible around \(t = 10\), which may arise from splitting the job into two separate submissions. (b) Post-processed results: \(\bullet\) markers show the original raw data, \(\square\) markers are the refined reference circuit results, and \(\blacktriangle\) markers represent the final corrected signal obtained by dividing the raw data by the refined reference.
}
  \label{fig:torino_gate_res}
\end{figure*}

Current quantum hardware remains highly noisy due to gate errors, decoherence, and readout imperfections. As a result, applying error mitigation techniques is needed~\cite{Shinjo_2024}. One strategy is to estimate gate-induced errors by preparing a reference circuit in which all gate parameters are set to zero. In this case, the time-evolution circuit becomes an identity operation, allowing one to measure the accumulated noise during the time evolution. In our implementation, however, we further refine this approach by exploiting the following commutation relation:
\begin{equation}
    \left[\sum_i{\sigma^z_{i}\sigma^z_{i+1}}, \sigma^z_{\text{cen}}\right]=0
\end{equation}
which implies that the observable $\langle \sigma^z_{\text{cen}}(t) \rangle$ is a conserved quantity. Based on this, we construct a reference circuit in which all \texttt{RX} and \texttt{RZ} gate angles are set to zero, while the \texttt{RZZ} gates remain unchanged. Figure~\ref{fig:torino_gate_res}(a) shows the results of our error mitigation procedure performed on  \texttt{ibm\_torino}, together with the original strategy with all gate parameters set to zero. 
As seen in the figure, the refined method (\(\square\) markers) captures the decay envelope of the dynamics more accurately, whereas the original method (\(\triangle\) markers) fails to do so. Figure~\ref{fig:torino_gate_res}(b) shows the final mitigated result, where the raw data (\(\bullet\) markers) is divided by the refined reference (\(\square\) markers); the corrected signal is shown as the \(\blacktriangle\) markers.

To closely track the evolving state of the quantum hardware, we construct the physical and reference circuits alternatively as the following order: $[U(t_0), U_{\text{ref}}(t_0), U(t_1), U_{\text{ref}}(t_1), \ldots]$, where \(U(t_i)\) denotes the circuit implementing the time evolution at step \(t_i\), and \(U_{\text{ref}}(t_i)\) is its corresponding reference circuit. The time steps are defined as \(t_i = i\, dt\), with \(dt = 0.1\). This alternative structure ensures that each reference circuit runs under  similar conditions to its corresponding evolution circuit. In our implementation, we manually set the number of circuits per job to 200, which results in the simulations for \(t \leq 10\) and \(t > 10\) being split across two separate jobs, which can account for  the discontinuity in Fig.~\ref{fig:torino_gate_res} that appears around $t = 10$.

\section{Riemannian quantum circuit optimization}
\label{sec:TN_Compress}

\begin{figure}[tbp]
  \centering
  \includegraphics[scale=0.5]{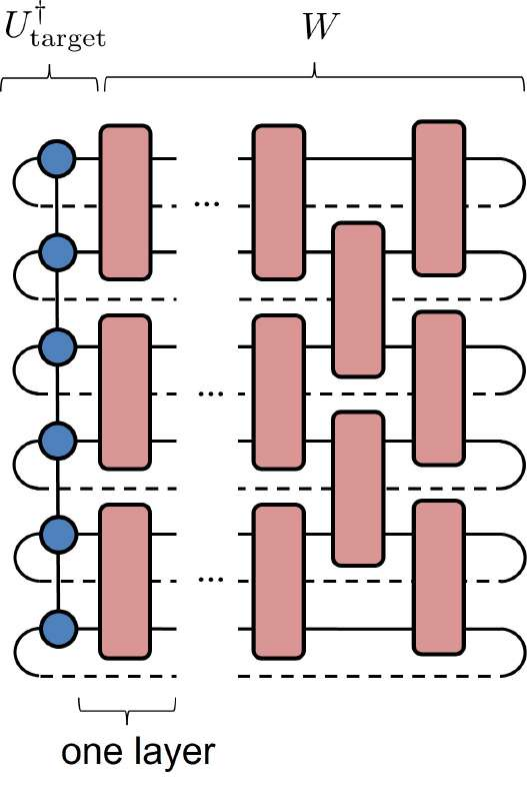}
  \caption{
TN representation of the overlap between the target unitary operator and the quantum gates. The blue tensors MPO form of the real-time evolution operator, while the pink tensors correspond to the quantum circuit with a fixed number of layers, initialized using second-order Trotter decomposition. This overlap corresponds to the trace term in the cost function defined in Eq.~\ref{eq:compress_cost}. The goal of the Riemannian optimization is to optimize the quantum gates such that this overlap is maximized, i.e., the cost function is minimized.
}
  \label{fig:qc_compress}
\end{figure}

Current quantum hardware remains limited by significant noise and decoherence, particularly for deep circuits. Therefore, reducing circuit depth is essential for improving simulation fidelity. To this end, we explore a compression technique called Riemannian quantum
circuit optimization~\cite{Le_2025,kotil2024riemannian,Putterer2025riemannian}. This approach constructs a low-depth quantum circuit to approximate a target unitary by treating the target unitary (the real-time evolution operator \(e^{-iHdt}\)) as a matrix product operator (MPO) and optimizing the quantum gates ( Fig.~\ref{fig:qc_compress}). The initial circuit is prepared as a second-order Trotterized evolution operator, and then TN techniques with Riemannian optimization are applied to search for the optimal gate sequence using a fixed number of layers. The optimization is performed on a classical computer, and once the optimal gates are obtained, the resulting circuit can be implemented on a quantum device. We can define the cost function
\begin{equation}\label{eq:compress_cost}
C_F(U_{\text{target}}, W) = 1 - \frac{1}{d}\text{Re}\left[ \text{Tr}\left(U_{\text{target}}^\dagger W \right) \right] ,
\end{equation}
which corresponds to the normalized Frobenius norm ~\cite{Le_2025}. Here, \(U_{\text{target}}\) is the target unitary represented as a MPO, and \(W\) is the quantum circuit composed of a fixed sequence of gates. The factor \(d = 2^N\), with \(N\) the number of qubits. The trace term can be computed using TN contraction, as illustrated in Fig.~\ref{fig:qc_compress}.

\begin{figure*}[tbp]
  \centering
  \includegraphics[width=\linewidth]{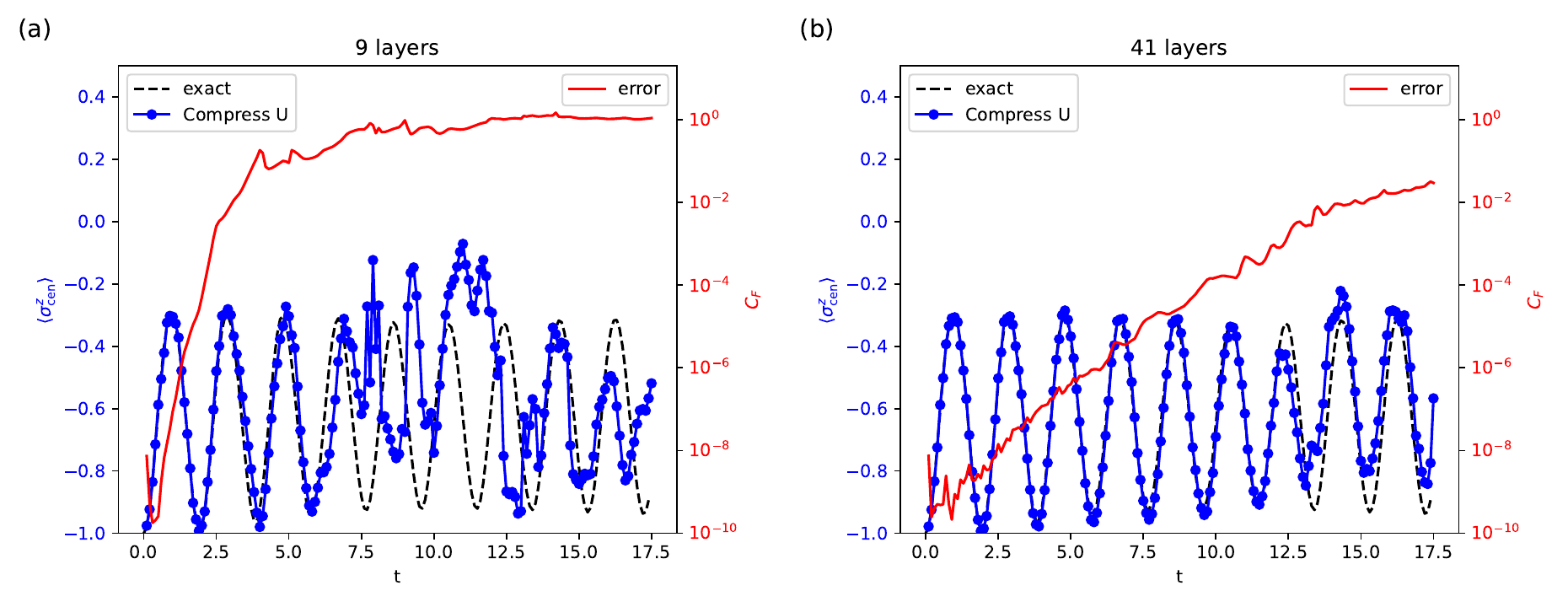}
  \caption{
Simulation results using Riemannian optimization to construct time-evolution circuits with different numbers of layers, executed on a classical simulator. The initial state is \(|\uparrow \uparrow \downarrow \downarrow \downarrow \downarrow \uparrow \uparrow\rangle\), and \(\langle \sigma^z(t) \rangle\) is measured at site 4. The model parameters are \(h_x = 1\), \(h_z = 3\), and each simulation uses 8192 shots. (a) Results using 9 circuit layers; (b) results using 41 layers. The black dashed line represents the exact evolution, the blue line with marker shows the simulator results with the optimized quantum gates, and the red solid line indicates the cost function defined in Eq.~\ref{eq:compress_cost}. In both cases, the fidelity of the simulation decreases over time, and longer-time evolution requires more circuit layers to maintain accuracy.
}
  \label{fig:diff_layer_opt}
\end{figure*}

To evaluate the performance of the Riemannian optimization approach, we first simulate real-time dynamics using classical simulator based on the optimized quantum gates prepared as shown in Fig.~\ref{fig:diff_layer_opt}. The quantum circuits are initialized using the optimized gates, and the resulting dynamics of $\langle \sigma^z_{\text{cen}}(t) \rangle$ are compared with the ED results. Additionally, we compute the cost function defined in Eq.~\eqref{eq:compress_cost} as an indicator of circuit accuracy. As expected, the approximation error accumulates over time, causing deviation from the exact evolution. The rate of this error growth depends on the depth of the circuit: circuits with more layers exhibit slower error accumulation and remain accurate for longer times.
This observation suggests a practical strategy for optimizing circuit resources: {adapt the number of layers based on the total desired evolution time}. That is, shallow circuits are adequate for short-time dynamics, while longer evolutions require deeper circuits to preserve fidelity. This layer-adaptive strategy helps reduce the number of quantum gates deployed on real hardware.

\begin{figure}[htbp]
  \centering
  \includegraphics[scale=0.6]{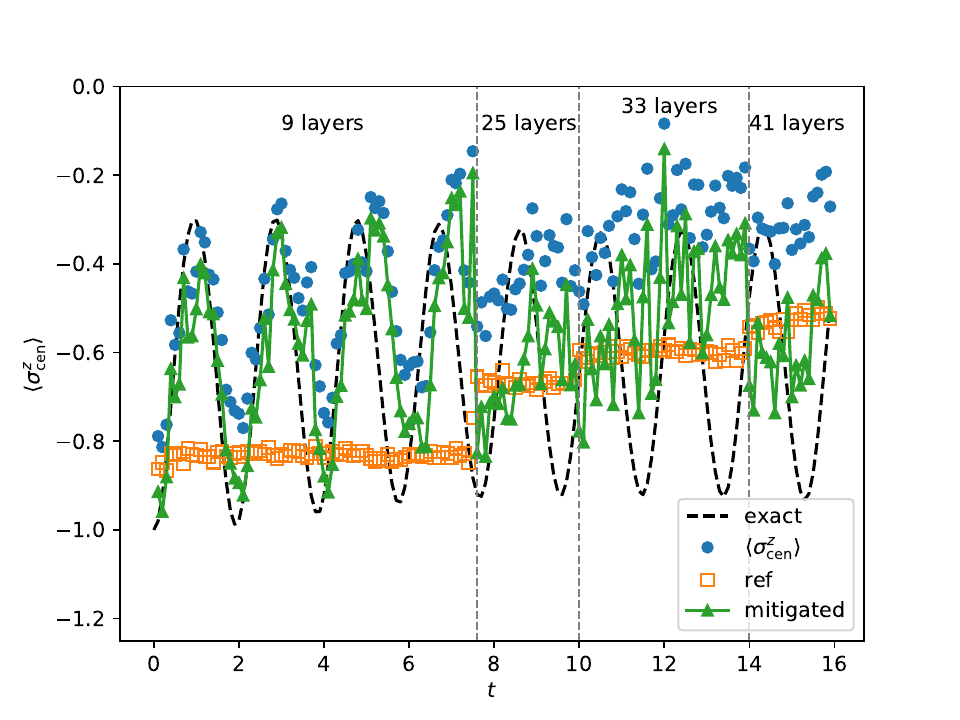}
  \caption{
Real-device simulation results using  \texttt{ibm\_torino}, where the time-evolution gates are prepared via Riemannian optimization. The model parameters and number of shots are the same as in Fig.~\ref{fig:diff_layer_opt}. The black dashed line shows the exact evolution of \(\langle \sigma^z(t) \rangle\) at site 4. \(\bullet\) markers represent the raw measurement results from the real device, \(\square\) markers correspond to the reference circuit in which all native gate parameters are set to zero after transpilation except for the \texttt{RZZ} gates, which remain unchanged, and \(\blacktriangle\) markers show the final error-mitigated result obtained by dividing the raw data by the reference. Vertical gray dashed lines indicate the time points where the number of circuit layers increases.
}
  \label{fig:ibm_torino_compress}
\end{figure}

In Fig.~\ref{fig:ibm_torino_compress}, we present results where the quantum gates obtained via Riemannian optimization are executed on  \texttt{ibm\_torino}, with the number of layers adjusted according to the evolution time. 
The central magnetization $\langle \sigma^z_{\text{cen}}(t) \rangle$ are obtained from the device. To mitigate hardware noise, we also run a reference circuit in which all parameters are set to zero except for the \texttt{RZZ} gates~\footnote{To perform the simulation on real quantum hardware, the unitary gates must be transpiled into native gates. In the reference circuit, all gate parameters are set to zero except for the \texttt{RZZ} gates, which remain unchanged.}. The result of this reference measurement is also shown in the figure. We observe that the signal from the reference circuit exhibits a step-like decay, with boundaries corresponding to points where the number of layers increases. Finally, we apply error mitigation by dividing the raw signal by the reference.

The simulation results on the real quantum device become very noisy when the number of layers exceeds 25 as shown in Fig.~\ref{fig:ibm_torino_compress}. This is due to the fact that transpilation of the optimized circuits  on real hardware results in deeper circuits of native gates. Therefore, if the circuit can be directly simulated using  native \texttt{RZ}, \texttt{RX}, and \texttt{RZZ} gates, the simulation can be less noisy, as we demonstrated previously.  Nonetheless, for more complicated Hamiltonians {when} the evolution operator can not be directly constructed from native gates, the Riemannian optimization compression will become useful.

\section{Results}
\label{sec:Freq_domain}
\begin{figure*}[tbp]
  \centering
  \includegraphics[width=\linewidth]{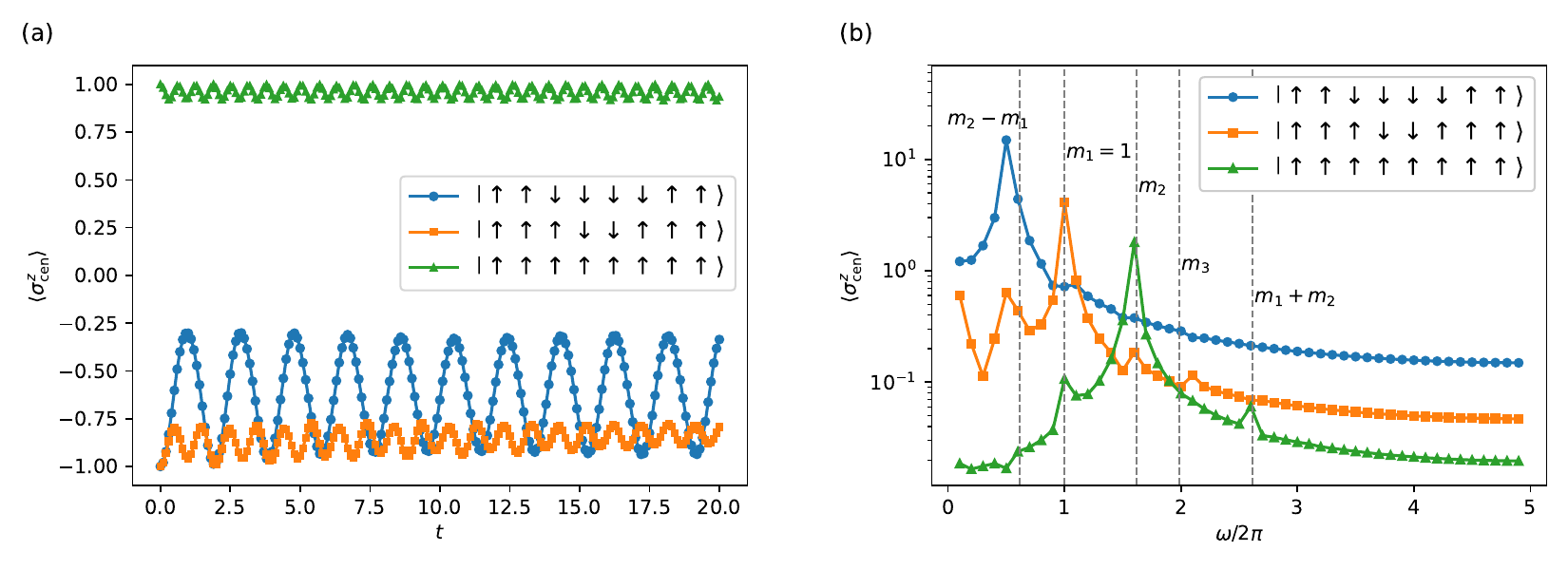}
  \caption{
Exact real-time evolution starting from different initial states for the transverse-field Ising model with parameters \( h_x = 1\) and \(h_z = 3\). (a) Time-domain results showing \(\langle \sigma^z_{\text{cen}}(t) \rangle\). (b) corresponding frequency-domain spectra obtained by Fourier transforming the data in (a), restricted to \(t \leq 10\). The gray dashed line indicate the expected position of peaks predict from Tab.~\ref{tab:e8_mass_spectrum}. We fix $m_1 = 1$. Different initial states produce distinct dominant frequency peaks, all of which align with the predicted $E_{8}$ mass spectrum.
}
  \label{fig:initial_diff_simulator}
\end{figure*}

Fourier transformation of the time evolution of $\langle \sigma^z_{\text{cen}}(t) \rangle$ allows us to examine the excitation spectrum to identify the masses of the excitations.  One challenge in this analysis is that, as shown in Fig.~\ref{fig:L11_E8}, only a few dominant peaks (typically one or two) appear in the frequency spectrum while the rest of the peaks {have very weak intensity}. The signal becomes increasingly noisy on real devices during the real-time evolution and it is difficult to extract those weaker {signals} after Fourier transformation.
To address this, we utilize the strategy of starting from different initial states, each potentially exciting different main frequencies in the spectrum. Fig.~\ref{fig:initial_diff_simulator}(a) shows the exact real-time evolution computed on a classical simulator for several different initial states. Each initial configuration leads to a distinct dominant frequency.
Recognizing that long-time quantum simulations are unreliable due to accumulated noise, we restrict the Fourier transform to data up to \(t = 10\). The resulting frequency-domain spectra are shown in Fig.~\ref{fig:initial_diff_simulator}(b). Even within this limited time window, distinct main frequency peaks emerge from different initial states. This approach makes it possible to access a broader range of the $E_8$ mass spectrum by combining the frequency responses from multiple  initial kink states.

\begin{table}[tbp]
\centering
\begin{tabular}{ccccccc c}
\toprule
\textbf{Label} &
\textbf{$E_8$ prediction} &
\multicolumn{2}{c}{\textbf{ED}} &
\multicolumn{2}{c}{\textbf{Trotter}} &
\multicolumn{2}{c}{\textbf{Riemannian}} \\
\cmidrule(lr){3-4}\cmidrule(lr){5-6}\cmidrule(lr){7-8}
 &  & Value & Deviation & Value & Deviation & Value & Deviation \\
\midrule
$m_{2}-m_{1}$ & $0.618\,m_{1}$ &
$0.5 \,m_{1}$ & $-0.118\,m_{1}$ &
$0.6\,m_{1}$ & $-0.018\,m_{1}$ &
$0.5\,m_{1}$ & $-0.118\,m_{1}$ \\

$m_{1}$       & $m_{1}$ &
$1.0\,m_{1}$ & $0$ &
$1.1\,m_{1}$ & $+0.10\,m_{1}$ &
$1.0\,m_{1}$ & $0$ \\

$m_{2}$       & $1.618\,m_{1}$ &
$1.6\,m_{1}$ & $-0.018\,m_{1}$ &
$1.7\,m_{1}$ & $+0.082\,m_{1}$ &
$1.6\,m_{1}$ & $-0.018\,m_{1}$ \\

$m_{1}+m_{2}$ & $2.618\,m_{1}$ &
$2.6\,m_{1}$ & $-0.018\,m_{1}$ &
$2.5\,m_{1}$ & $-0.118\,m_{1}$ &
$2.6\,m_{1}$ & $-0.018\,m_{1}$ \\
\bottomrule
\end{tabular}
\caption{
{The first two columns list the} first few mass spectrum predicted by the $E_8$ field theory. Each mass values of $m_i$ is expressed in terms of the lightest mass $m_1$. 
The next column pair reports the masses obtained from ED, while the following two column pairs present the masses extracted from the IBM Quantum device \texttt{ibm\_torino} using first-order Trotter decomposition and Riemannian-optimized circuits, respectively.  
Deviations are computed as the difference between the measured value and the $E_8$ prediction.   
The corresponding peak positions in the frequency spectra are those highlighted with circled markers in Fig.~\ref{fig:initial_diff_real_device}.
See text for more discussion.
}

\label{tab:e8_mass_spectrum}
\end{table}

\begin{figure*}[tbp]
  \centering
  \includegraphics[width=\linewidth]{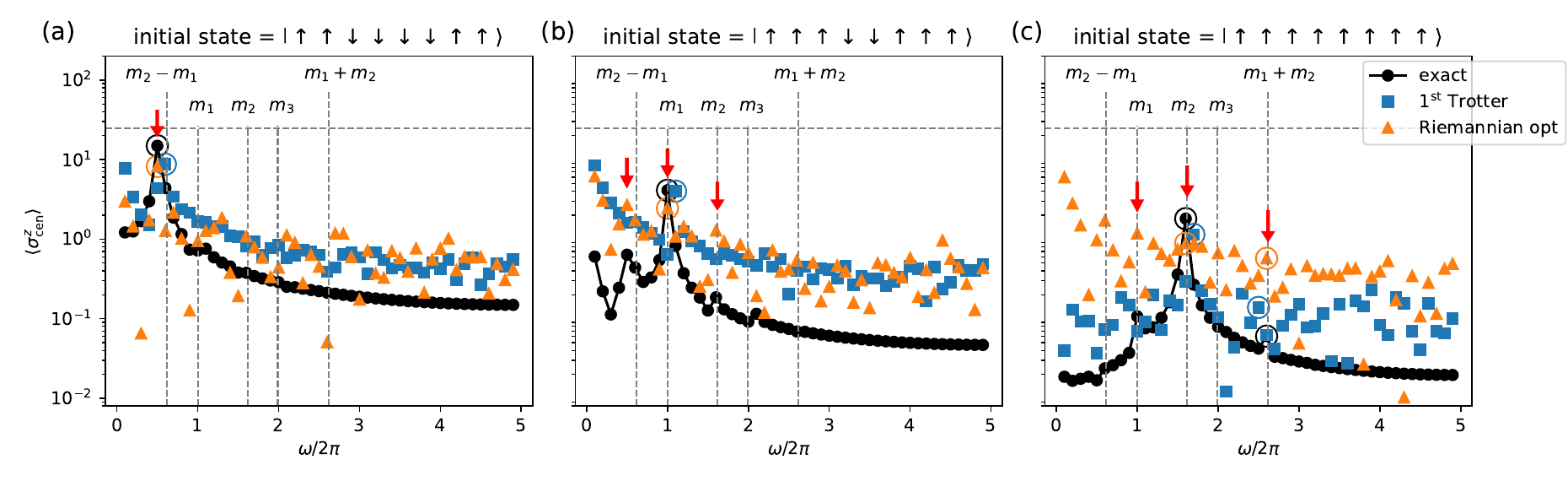}
  \caption{
Frequency-domain simulation results on the IBM Quantum device \texttt{ibm\_torino}, with model parameters and initial states matching those in Fig.~\ref{fig:initial_diff_simulator}, and using 8192 shots per circuit. \(\bullet\) markers show the exact result from the classical computer, \(\blacksquare\) markers correspond to results obtained by constructing the evolution gates using first-order Trotterization, and \(\blacktriangle\) markers represent results using gates prepared via Riemannian optimization. Vertical gray dashed lines indicate the expected peak positions predicted by the $E_{8}$ spectrum, with \(m_i\) denoting the expected meson masses. Red arrows mark the peaks observed both in the exact simulation and in the quantum-device results. The circled points correspond to the peak values listed in Tab.~\ref{tab:e8_mass_spectrum}. Both quantum-device methods include error mitigation before applying the Fourier transform to extract the spectra. All subfigures use data up to \(t = 10\).
}
  \label{fig:initial_diff_real_device}
\end{figure*}

We also perform real-time evolution on a real quantum device, starting from various initial states to explore their frequency responses.  Figure~\ref{fig:initial_diff_real_device} shows the results from the two approaches.
For both methods, error mitigation is applied by setting all native gate parameters to zero except for the \texttt{RZZ} gates. After collecting measurement data from the real device and {performing} the error mitigation, we perform  Fourier transform of the time series to the frequency-domain. We find that each initial state yields a clear dominant peak, and in some cases, even a secondary peak is visible. From these results, we can observe at least the peaks corresponding to \(m_2 - m_1\), \(m_1\), and \(m_2\) across different initial states. 
In Fig.~\ref{fig:initial_diff_real_device}, the spectral points corresponding to these identified peaks are highlighted with circles, and their extracted values are summarized in Tab.~\ref{tab:e8_mass_spectrum}, including those from  ED, Trotterized and Riemannian-optimized circuits. The table lists the measured peak positions together with their deviation relative to the expected $E_8$ values. Most of the extracted peaks lie within the resolution limit of the frequency \(d\omega = 0.1\times 2\pi\)~\footnote{Because the real-device evolution is performed only up to \(t_{\max}=10\) with a time step \(d t = 0.1\), the frequency resolution is limited to \(d\omega = 2\pi/t_{\max} = 0.1\times 2\pi\).}.  
We also note that among the extracted features, both the \(m_1+m_2\) and \(m_2-m_1\) peaks exhibit deviations larger than \(d\omega\).  
We note that although direct Trotterization using native gates gives a cleaner time series, the Riemannian optimization gives consistent results with ED amid lower signal-to-noise ratio.

\section{Conclusion}
\label{sec:conclusion}

We demonstrate the  state-of-the-art simulations of $E_8$ mass {spectra}  using the IBM  Quantum device \texttt{ibm\_torino}, which  supports native \texttt{RX}, \texttt{RZ} and \texttt{RZZ} gates. 
By benchmarking two approaches of the real-time evolution, we find direct construction of the Trotterized evolution unitary using  native gates and the Riemannian compression of the evolution operator into a fixed-layer quantum circuit gives similar performance.
However, we expect the Riemannian optimization should perform better when the time-evolution operators can not be expressed in native gates. 
Although simulating real-time evolution on a gate-based digital quantum computer suffers from the Trotterization error, our results are comparable to those obtained through the analog Ryderberg atom system~\cite{Vovrosh_2025}. This demonstrates that, despite the current noise limitations in quantum hardware, essential physical features, such as  low-lying mass peaks in the $E_{8}$ spectrum, can be successfully observed and resolved. These results show that current quantum devices, when combined with classical techniques like circuit compression and error mitigation, can be used to explore nontrivial quantum field theory phenomena, and with limited circuit resources, physically meaningful information such as mass spectra can still be extracted.

Looking forward, an exciting direction is to extend these techniques to the quantum quench dynamics of the XXZ spin chain. Although preparing its arbitrary ground state is challenging on current hardware~\cite{Van_Dyke_2022}, simple product states such as the N\'eel state or dimer product state can be readily initialized, enabling quench protocols similar to those demonstrated in ~\cite{Smith_2019}. With optimized low-depth circuits and improved error mitigation, real devices may access sufficiently long evolution times to allow comparison with theoretical predictions from the overlap-thermodynamic Bethe ansatz (OTBA)~\cite{Pozsgay_2014}. Such an approach would provide a promising route toward experimentally exploring out-of-equilibrium dynamics in integrable models using quantum processors.

A further potential direction is the massive Thirring model, whose lattice discretization can be mapped to the XXZ spin chain coupled to both uniform and staggered magnetic fields via the Jordan-Wigner transformation~\cite{MariCarmen_2019}. Recent numerical studies have explored its nonequilibrium dynamics under quantum quenches~\cite{MariCarmen_2025}. Implementing such quenches on gate-based quantum hardware would open a complementary route to studying real-time dynamics in quantum field theories. Extending these methods to both spin-chain and field-theory models would expand the out-of-equilibrium physics accessible on near-term devices.

\ack{We thank Rongyang Sun, Tzu-Chieh Wei for helpful discussions. We also acknowledge  IBM Q Hub at NTU for providing access to quantum hardware and computational resources used in the real-device simulations.}

\funding{H.-T.H. and  Y.-J.K. acknowledge support from the NSTC of Taiwan under Grant No. 113-2112-M-002-033-MY3.  This research was supported in part by grant NSF PHY-2309135 to the Kavli Institute for Theoretical Physics (KITP).}

\data{The data and code used to generate the figures in this work are available on GitHub~\url{https://github.com/hunghaoti/E8_QC_data}.
}

\appendix
\renewcommand{\thesection}{Appendix~\Alph{section}}
\section{Simulation results using Exact Diagonalization and Tensor Network}
\label{appdx:TN_res}
\begin{figure*}[htbp]
  \includegraphics[width=\linewidth, trim = 0 0 0 0]
  {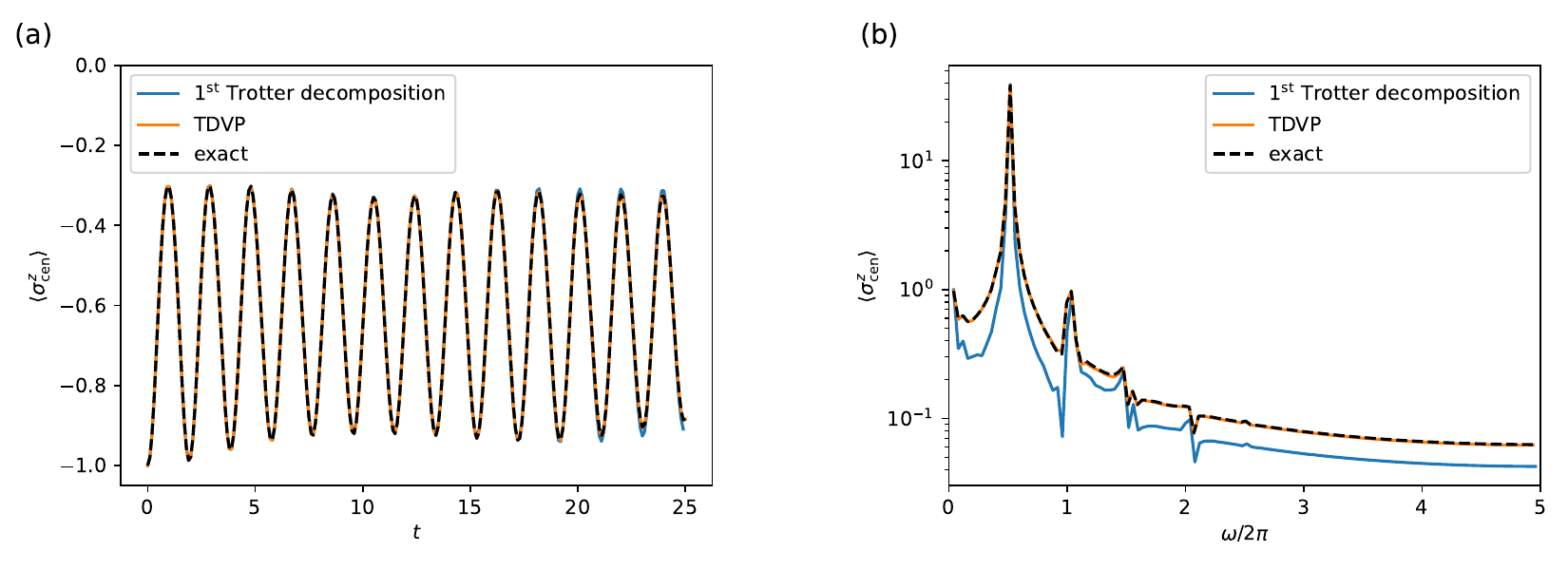}
  \caption{
Classical simulation of the real-time evolution of the central-site magnetization \(\langle \sigma^z_{\text{cen}}(t) \rangle\) using three methods: exact diagonalization (ED), time-dependent variational principle (TDVP), and first-order Trotter decomposition. Parameters are set to \(h_x = 1\), \(h_z = 3\), {with system size \(L = 8\) and initial state \(|\!\uparrow \uparrow \downarrow \downarrow \downarrow \downarrow \uparrow \uparrow\rangle\)}. The bond dimension of TDVP is $D=8$. (a) shows the results in the time domain, and (b) shows the corresponding frequency-domain spectrum obtained via Fourier transform. 
}
  \label{fig:tdvp_trotter}
\end{figure*}

To validate the quantum simulation results presented in the main text, we also perform classical simulations of the real-time dynamics using multiple methods:

\begin{itemize}
    \item[(1)] Exact Diagonalization (ED, as shown in Fig.\ref{fig:L11_E8} in the main text).
    \item[(2)] Time-Dependent Variational Principle (TDVP).
    \item[(3)] First-Order Trotter Decomposition via TN. This approach constructs the time-evolution operator as a product of local gates using first-order Trotterization and evolves the state using TN techniques. It mimics the statevector dynamics of quantum circuits on classical hardware.
\end{itemize}
The simulations are implemented using the \texttt{Cytnx} library~\cite{Cytnx_2025}.
All simulations are performed for the Hamiltonian defined in Eq.~\eqref{eq:Ising_H}, with parameters set to \(h_x = 1\), \(h_z = 3\), and system size \(L = 8\). The initial state is chosen as \(|\!\uparrow \uparrow \downarrow \downarrow \downarrow \downarrow \uparrow \uparrow\rangle\), and we track the time evolution of the central-site magnetization \(\langle \sigma^z_{\text{cen}}(t) \rangle\).
Fig.~\ref{fig:tdvp_trotter}(a) shows the time-domain results from all three methods. The TDVP simulation closely agrees with the ED with bond dimension only set as \(D=8\). The first-order Trotter decomposition also performs well for short to intermediate times, but small deviations emerge at longer times due to accumulated Trotter errors. While these discrepancies slightly distort the frequency-domain spectrum, the dominant peak positions remain consistent, as seen in the Fourier-transformed data \(\langle \sigma^z_{\text{cen}}(\omega) \rangle\) (see Fig.~\ref{fig:tdvp_trotter}(b)). Importantly, these small Trotter errors are negligible compared to the level of noise encountered in quantum real-device simulations.

\begin{figure}[htbp]
  \includegraphics[scale=0.6]
  {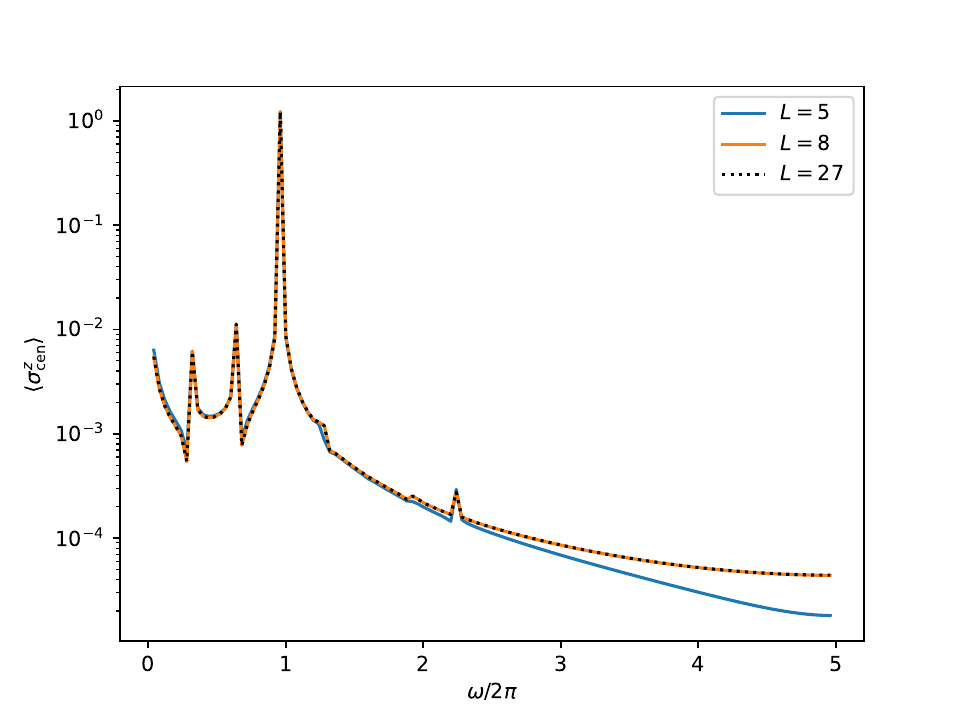}
  \centering
  \caption{
Frequency-domain spectrum \(\langle \sigma^z_{\text{cen}}(\omega) \rangle\) obtained via Trotter decomposition for different system sizes: \(L = 5\), \(8\), and \(27\). Initial states are prepared such that the two central sites are in the down state while all other sites are in the up state. For the even-sized system (\(L = 8\)), the central-left site (site 4) is used for measurement.
}

  \label{fig:diff_sites}
\end{figure}

To explore finite-size effects, we further conduct simulations for different system sizes \(L = 5\), \(8\), and \(27\). The time evolution is simulated up to \(t = 25\) with a time step \(d t = 0.1\). The resulting spectrum in the frequency domain is shown in Fig.~\ref{fig:diff_sites}. The results indicate that \(L = 8\) is already sufficient to capture the key spectral features of interest. This suggests that relatively small system sizes are sufficient to extract meaningful information about the $E_8$ mass spectrum in this model.

\section{IBM Quantum Device Settings }
\label{appdx:qc_set}
In this appendix, we summarize the technical details of how the quantum simulations were executed on IBM Quantum hardware, specifically on the \texttt{ibm\_torino} backend. This device is part of IBM’s Heron-class QPUs, which support fractional gates.

One critical setting in our experiments is enabling native fractional gates during backend selection. This allows us to access the \texttt{RZZ} gate directly as a native hardware-supported operation, which effectively reduces the circuit depth by avoiding additional gate decomposition. In Qiskit, this is done by initializing the backend with the \texttt{use\_fractional\_gates=True} option.
The use of fractional gates is currently supported only on Heron processors, such as \texttt{ibm\_torino}, and is essential for expressing two-qubit gate like \texttt{RZZ} natively.
Additionally, when transpiling our circuits to match the native gate set of the backend, we set the transpiler \texttt{optimization\_level=0} using Qiskit’s \texttt{generate\_preset\_pass\_manager}. We find that if we set \texttt{optimization\_level=2} leads to error where the measured observable becomes nearly constant during the entire evolution.

\begin{figure}[htbp]
  \includegraphics[scale=0.6]{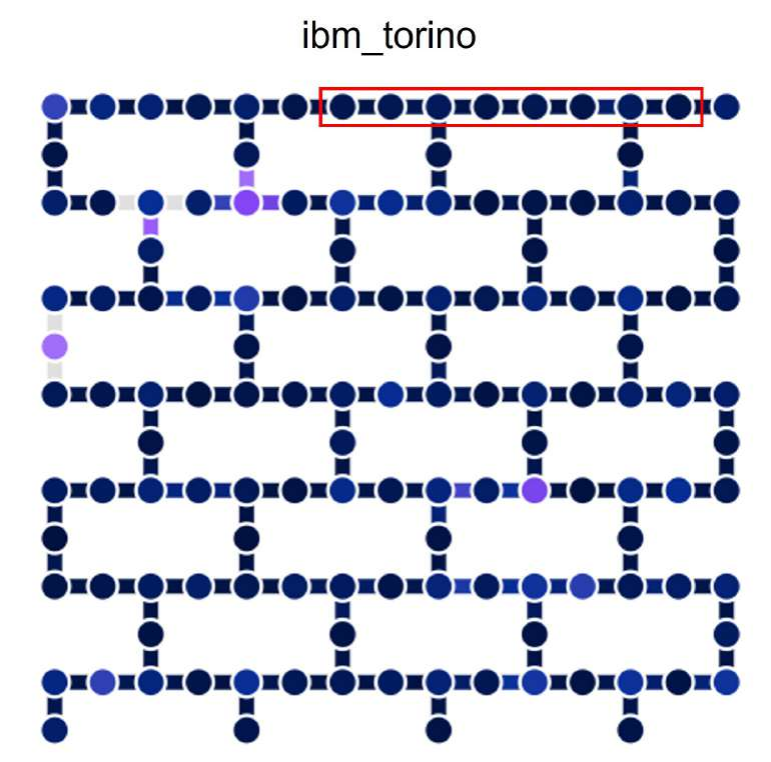}
  \centering
  \caption{Qubit layout of the IBM Quantum device \texttt{ibm\_torino}, a Heron-class processor used in our experiments. The red rectangle contains the QPUs selected for our simulations.
  }
  \label{fig:ibm_torino}
\end{figure}

In our simulations, we manually select the specifically qubits to ensure better connectivity and reduced noise. The layout of the quantum processor used in \texttt{ibm\_torino}, including the chosen qubits highlighted in red, is shown in Fig.~\ref{fig:ibm_torino}.

\bibliographystyle{iopart-num}
\bibliography{refs} 

\end{document}